# Microwave Response of Ceramic MgB$_2$ Samples

A. Agliolo Gallitto, G. Bonsignore, S. Fricano, M. Guccione and M. Li Vigni

*INFM and Dipartimento di Scienze Fisiche e Astronomiche, Università di Palermo, via Archirafi 36, 90123 Palermo, Italy*

**Abstract**

The microwave response of ceramic MgB$_2$ has been investigated as a function of temperature and external magnetic field by two different techniques: microwave surface impedance and second-harmonic emission measurements. The measurements of the surface resistance have shown that microwave losses in MgB$_2$ are strongly affected by the magnetic field in the whole range of temperatures below $T_c$, even for relatively low field values. The results have been accounted for in the framework of the Coffey and Clem model hypothesizing that in different temperature ranges the microwave current induces fluxons to move in different regimes. In particular, the results at temperatures close to $T_c$ have been quantitatively justified by assuming that fluxons move in the flux-flow regime and taking into account the anisotropy of the upper critical field. At low temperatures, the field dependence of the surface resistance follows the law expected in the pinning limit; however, an unusually enhanced field variation has been detected, which could be due to the peculiar fluxon structure of MgB$_2$, related to the presence of the two gaps. The measurements of the second-harmonic signals have highlighted several mechanisms responsible for the nonlinear response. At low magnetic fields and low temperatures, the nonlinear response is due to processes involving weak links. At temperatures close to $T_c$, a further contribution to the harmonic emission is present; it arises from the modulation of the order parameter by the microwave field and gives rise to a peak in the temperature dependence of the harmonic-signal intensity.

**Corresponding Author**

Aurelio Agliolo Gallitto
Dipartimento di Scienze Fisiche e Astronomiche, Università di Palermo
Via Archirafi 36, I-90123 Palermo (Italy)
Tel +39.091.6234207 - Fax +39.091.6162461
E-mail: agliolo@fisica.unipa.it - http://www.fisica.unipa.it

# 1. Introduction

The discovery of the superconductivity at 40 K in magnesium diboride [1] has driven the interest of the scientific community toward the investigation of the properties of this compound in both the normal and superconducting states [2-16]. A large number of studies have been devoted to understanding the coupling mechanism responsible for superconductivity in $MgB_2$ as well as to explore its potential for technological applications [17, 18]. Indeed, the malleability and ductility of the compound, due to its metallic nature, make $MgB_2$ a promising candidate for realizing wires, tapes and layers even of large areas, to be used for a large variety of devices.

One of the technological fields in which the use of superconducting materials is particularly convenient concerns the implementation of superconductor-based microwave devices [19, 20]. The study of the microwave response of superconductors allows determining specific properties of the investigated samples, of great value for their technological applications, as well as measuring basic properties of the superconducting state [8, 18, 21, 22]. This investigation is usually performed by measuring the surface impedance [8-10, 18, 21-24] or, alternatively, the power absorption with EPR spectrometers [25, 26]. A further method concerns the study of the harmonic emission [27-35]; in this case, the samples are exposed to an intense microwave field oscillating at the fundamental frequency $\omega$ and the signals radiated at the harmonic frequencies of the driving field are detected. Measurements of harmonic emission allow investigating the mechanisms by which part of the absorbed power conveys into harmonic signals, yielding complementary information to the direct power absorption method.

In this paper, we report experimental results on the microwave response of ceramic $MgB_2$ samples, produced by different methods. The microwave response has been investigated by means of two different techniques: surface impedance and harmonic emission measurements. The microwave surface resistance and the signal radiated at the second-harmonic (SH) frequency of the driving field have been investigated as a function of temperature and external magnetic field. The surface impedance measurements have shown that in $MgB_2$ the field-induced variations of the microwave surface resistance are unusually enhanced even at low temperatures, where the pinning effects should hinder the microwave losses. The SH measurements have highlighted that different mechanisms are responsible for the nonlinear response, whose effectiveness depends on temperature and intensity of the external magnetic field. In particular, our results show that the main source of harmonic emission at low magnetic fields and low temperatures is related to the presence of weak links.

The paper is organized in two parts. In Section 2, we report and discuss the results of microwave surface impedance. Since these measurements have been performed at low input power levels, the results concern the mw response in the linear regime. In Section 3, we discuss the results of harmonic emission. The experimental results, in both the linear and nonlinear regimes, are discussed in the framework of models reported in the literature. Particular attention will be devoted to the comparison between the microwave properties of $MgB_2$ and those of other superconductors.

## 2. Microwave Surface Impedance

The complex surface impedance, $Z_S = R_S - iX_S$, accounts for the absorption and reflection of high-frequency em waves at the surface of conductors. Measurements of both the resistive, $R_S$, and reactive, $X_S$, components of $Z_S$ in superconductors allow investigating important properties of the superconducting state, such as the gap parameter, quasiparticle density and nature of scattering [21]. Measurements of $R_S$ yield information about dissipative processes, while those of $X_S$ provide a convenient method for determining the temperature dependence of the field penetration depth $\lambda(T)$.

In this Section, we investigate the field-induced variations of the microwave surface resistance. It is well know that $R_S$ of superconductors in the mixed state depends on the magnetic field through several mechanisms [22, 24]. The different vortex states, in the different regions of the H-T plane, determine the temperature and field dependencies of $R_S$. Therefore, measurements of $R_S(H, T)$ provide important information on the fluxon dynamics in the different regimes of motion.

### *2.1 Experimental Apparatus and Samples*

The microwave surface resistance has been investigated in two different $MgB_2$ samples. One (P#1) consists of 10 mg of Alfa-Aesar powder. The other (P#2) has been obtained by milling a bulk piece of ceramic $MgB_2$, which had been synthesized by the direct reaction of boron powder with a lump of magnesium metal, both of purity better than 99.95% [37]. The powdering process of the P#2 sample has been performed in order to increase its effective surface, improving the measurement sensitivity, and make clearer the conditions for a comparison of the results in samples of similar shape. The two samples have equal weight and are both sealed in Plexiglas holders. The critical temperature is ≈ 39 K for both samples. From measurements performed in a bulk specimen of the same batch of P#2, the normal-state resistivity and the residual surface resistance at $T = 0$ have been determined: $\rho(T_c) \approx 15$ μΩcm and $R_{res}(0) \approx 6$ mΩ [8].

The experiments have been performed using the cavity perturbation technique. The cavity, of cylindrical shape with golden-plated walls, was tuned in the $TE_{011}$ mode resonating at 9.6 GHz. Its $Q$-factor is ~ 25000 at room temperature and ~ 40000 at the helium temperature. The sample is located at the bottom of the cavity, in a region where the microwave magnetic field is maximal, and placed between the expansions of an electromagnet, which generates fields up to ~ 10 kOe. Two additional coils, externally fed, allow reducing to zero the residual field of the electromagnet and working at low magnetic fields. The $Q$-factor of the cavity is measured by means of an *hp*-8719D Network Analyzer.

The surface resistance of the sample is proportional to $(1/Q_L - 1/Q_U)$, where $Q_U$ is the $Q$-factor of the empty cavity and $Q_L$ is that of the loaded cavity. In order to disregard the geometric factor of the sample, it is convenient to normalise the deduced values of $R_S$ to the value of the surface resistance at a fixed temperature in the normal state, $R_N$. Measurements have been performed as a function of the temperature and the static field, $H_0$. All the results have been obtained with the static magnetic field parallel to the microwave magnetic field. The input power is of the order of nW.

*2.2 Experimental Results*

Fig. 1 shows $(1/Q_L - 1/Q_U)$ as a function of the temperature in the absence of external magnetic fields, for both samples. As one can see, in the normal state the P#1 sample induces microwave losses five times greater than P#2; whereas, at low temperatures the microwave losses induced by the two samples differ each other by a factor of about two.

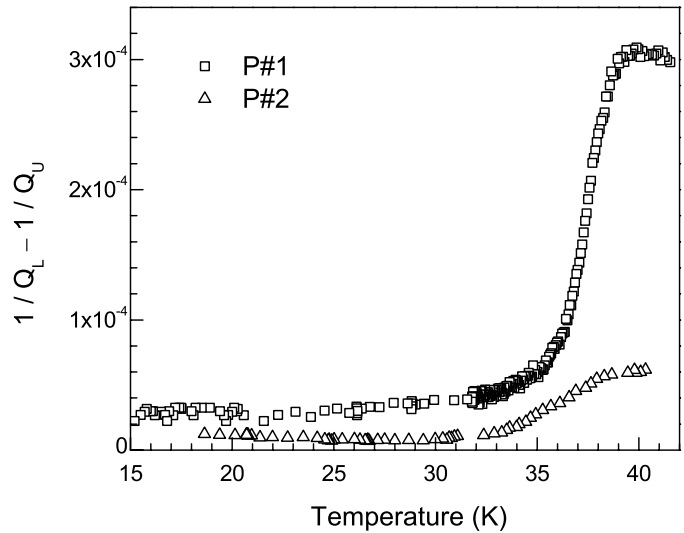

**Figure 1** – Temperature dependence of $(1/Q_L - 1/Q_U)$, for P#1 and P#2 samples, in the absence of static magnetic fields.

In Fig. 2 we report the temperature dependence of $R_S/R_N$ of P#1 and P#2, at different values of the static field. The results have been obtained according to the following procedure: the sample was zero-field cooled down to 4.2 K, then $H_0$ was set at a given value, which was kept constant during the time in which all measurements in the temperature range 4.2 ÷ 40 K have been performed. $R_S$ has been normalized to its value at $H_0 = 0$ and $T = 40$ K. Similarly to what happens in high-$T_c$ superconductors, on increasing $H_0$ the $R_S(T)$ curve broadens and shifts toward lower temperatures. However, the effect of the magnetic field on $R_S(T)$ in MgB$_2$ is more significant than that reported in the literature for cuprate superconductors [22-24]. In particular, one can see that a noticeable field variation of $R_S$ is observed even at the lowest temperatures, where the pinning effects would hinder the energy losses.

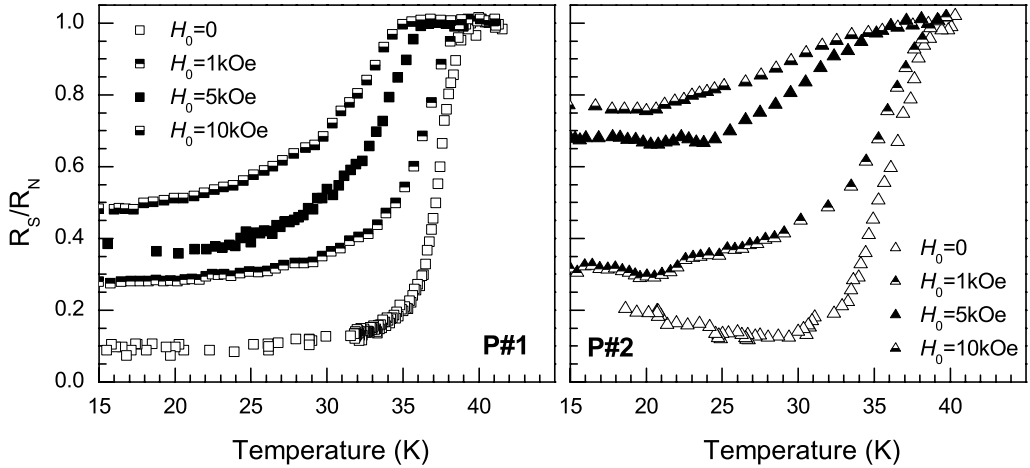

**Figure 2** – Normalized values of the microwave surface resistance of P#1 and P#2 samples as a function of the temperature, at different values of the static field. The normal-state surface resistance, $R_N$, has been determined at $H_0 = 0$ and $T = 40$ K.

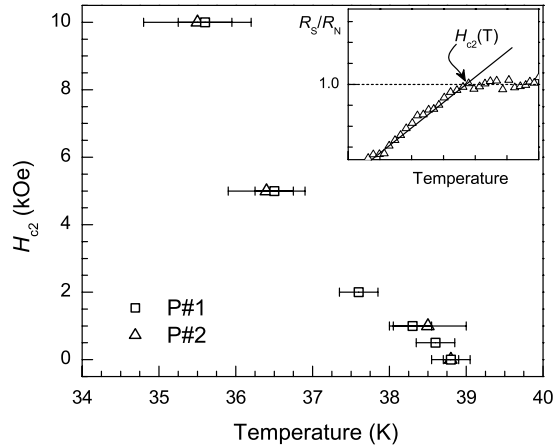

**Figure 3** – Temperature dependence of the upper critical field deduced from the $R_S(T, H_0)/R_N$ values, for both P#1 (squares) and P#2 (triangles) samples. The inset shows the way by which we have deduced $H_{c2}(T)$.

By measuring the shift of the transition temperature induced by $H_0$ we have deduced the temperature dependence of the upper critical field near $T_c$. The deduced values of $H_{c2}(T)$ are shown in Fig. 3, for both P#1 (squares) and P#2 (triangles) samples. The inset shows the way by which we have determined $H_{c2}(T)$. Both the $H_{c2}(T)$ curves show values of $(dH_{c2}/dT)_{T=T_c}$ of about 2 kOe/K and exhibit an upward curvature, consistent with that reported in the literature for $MgB_2$ [2-5].

Fig. 4 shows the normalized surface resistance as a function of the magnetic field at $T = 4.2$ K, for the two samples. Symbols are the experimental data, lines are the best-fit curves obtained with the procedure described in the next section.

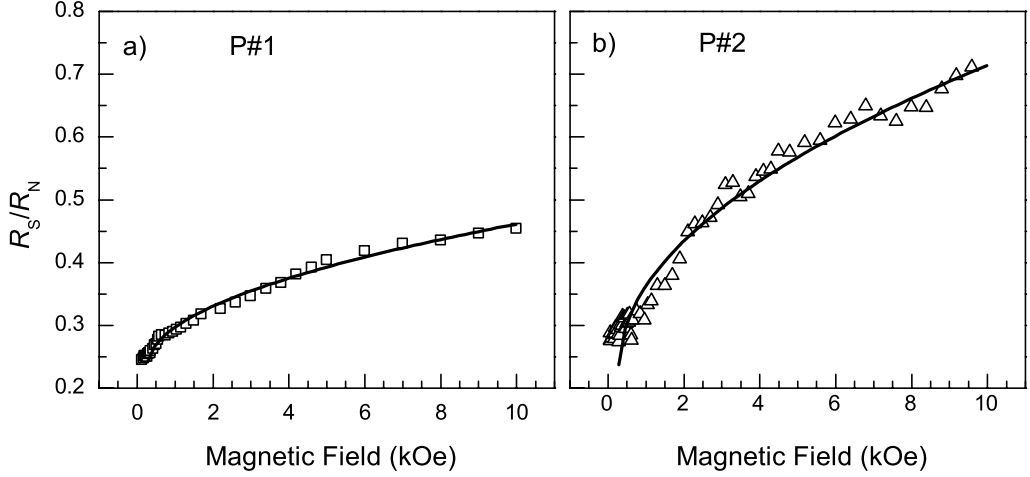

**Figure 4** – Magnetic field dependence of $R_S/R_N$ for the two samples of $MgB_2$, at $T = 6$ K. The lines are the best-fit curves of the experimental data, obtained using Eq. (2.9). The best-fit parameters are $H^* = 280$ Oe for both samples and $\alpha = 2.5 \times 10^{-3}$ for P#1 and $\alpha = 5 \times 10^{-3}$ for P#2.

Figs. 5 (a) and (b) show the normalized values of the surface resistance as a function of the static magnetic field for P#1 and P#2, respectively, at different values of the temperature, near $T_c$. As expected, on increasing the temperature the samples go into the normal state at lower values of the magnetic field. A comparison between panels (a) and (b) shows that the values of $R_S(T)/R_N$ of P#1 are smaller than those of P#2. Furthermore, although the two samples have approximately the same values of $H_{c2}(T)$, on increasing $H_0$, the surface resistance of P#2 approaches the normal-state value faster than that of P#1.

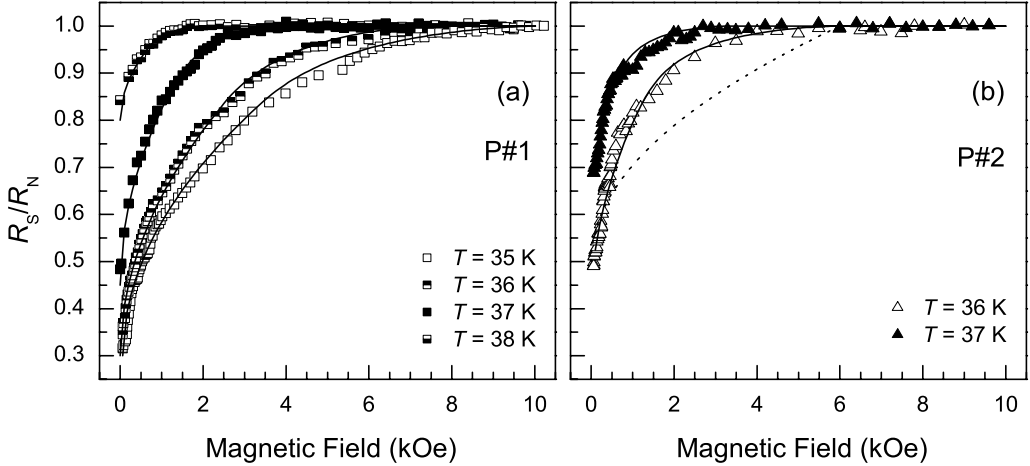

**Figure 5** – Normalized values of the surface resistance as a function of the static magnetic field for P#1 (a) and P#2 (b), at different values of the temperature, near $T_c$. Symbols are experimental data. Continuous lines are the best-fit curves, obtained by taking into account the upper-critical-field anisotropy as explained in the text. The lines in (a) have been obtained with $\lambda_0/\delta_0 = 10^{-2}$, $\gamma = 3$, $H_{c2}^{\perp c}(35) = 10$ kOe, $H_{c2}^{\perp c}(36) = 7.5$ kOe, $H_{c2}^{\perp c}(37) = 3.8$ kOe, $H_{c2}^{\perp c}(38) = 1.7$ kOe. The continuous lines in (b) have been obtained with $\lambda_0/\delta_0 = 10^{-2}$, $\gamma = 5$, $H_{c2}^{\perp c}(36) = 6$ kOe, $H_{c2}^{\perp c}(37) = 3.3$ kOe. The dashed line has been obtained with $\lambda_0/\delta_0 = 10^{-2}$ and $H_{c2}(36) = 6$ kOe, disregarding the anisotropy of the upper critical field.

*2.3 Discussion*

Microwave losses induced by static magnetic fields in high-$T_c$ superconductors have been investigated by several authors [22-25, 38-40]; they have been mainly ascribed to the motion of fluxons in the flux-flow regime. However, it has been pointed out that a noticeable contribution can arise from the presence of the normal fluid [23, 38], especially at temperatures close to $T_c$ and for magnetic fields of the same order of the upper critical field. Our results show that the effect of the applied magnetic field on the microwave surface resistance of $MgB_2$ is stronger than that observed in both conventional and cuprate superconductors, in all the range of temperatures $T < T_c$. In particular, from Fig. 4 one can see that an unusually enhanced field variation of $R_S$ is observed even at the lowest temperatures: $R_S$ takes on the values 0.45 $R_N$ and 0.7 $R_N$ for P#1 and P#2, respectively, when a magnetic field of just 10 kOe is applied. This property allows investigating the role of the magnetic field on the fluxon dynamics even at low temperatures, where the pinning effects would hinder the energy losses.

In the London local limit, the surface impedance is proportional to the complex penetration depth of the *em* field, $\tilde{\lambda}$. In particular

$$R_S = -\frac{4\pi\omega}{c^2}\mathrm{Im}(\tilde{\lambda}). \tag{2.1}$$

The complex penetration depth has been calculated in different approximations [38-40]. Coffey and Clem [39] have developed a comprehensive theory for the electromagnetic response of type-II superconductors in the mixed state, in the framework of the two-fluid model of the superconductivity. The theory applies for $H_0 > 2H_{c1}$, when the induction field inside the sample can be supposed as generated by an uniform density of fluxons; in this case $H_0 \approx B_0 = n\phi_0$, where $\phi_0$ is the flux quantum and $n$ is the vortex density. In their model, the authors calculate the complex penetration depth, $\tilde{\lambda}$, by taking into account the effects of the fluxon motion and the very presence of vortices, which bring along normal material in their cores. In the linear approximation, $H(\omega) \ll H_0$, they found the following expression for $\tilde{\lambda}$ [39]:

$$\tilde{\lambda}^2 = \frac{\lambda^2 + \tilde{\delta}_v^2}{1 - 2i\lambda^2/\delta^2} \tag{2.2}$$

where $\tilde{\delta}_v$ is the complex effective skin depth arising from vortex motion; $\lambda$ and $\delta$, the London penetration depth and the normal-fluid skin depth, are given by

$$\lambda = \frac{\lambda_0}{\sqrt{(1-w)(1-B_0/B_{C2})}} \qquad \delta = \frac{\delta_0}{\sqrt{1-(1-w)(1-B_0/B_{C2})}}; \tag{2.3}$$

here, $\lambda_0$ is the London penetration depth at $T = 0$; $\delta_0$ is the normal skin depth at $T = T_c$; $w$ is the fractions of normal electrons at $H_0 = 0$, in the two-fluid model $w = (T/T_c)^4$.

The penetration depth $\tilde{\delta}_v$ can be written in terms of two characteristic lengths, $\delta_f$ and $\lambda_c$, arising from the contributions of the viscous and the restoring-pinning forces, respectively:

$$\frac{1}{\tilde{\delta}_v^2} = \frac{1}{\lambda_c^2} - \frac{2i}{\delta_f^2} \tag{2.4}$$

where

$$\lambda_c^2 = \frac{B_0 \phi_0}{4\pi k_p} \tag{2.5}$$

$$\delta_f^2 = \frac{B_0 \phi_0}{2\pi \omega \eta} \tag{2.6}$$

with $k_p$ the restoring-force coefficient and $\eta$ the viscous-drag coefficient [41].
The effectiveness of the two terms in Eq. (2.4) depends on the ratio $\omega_c = k_p/\eta$, which defines the depinning frequency. In particular, when the frequency of the *em* wave, $\omega$, is much larger than $\omega_c$ the contribution of the viscous-drag force predominates and the induced *em* current makes fluxons moving in the flux-flow regime. On the contrary, for $\omega \ll \omega_c$ the motion of fluxons is ruled by the restoring-pinning force.

In order to discuss the experimental results we have to consider that in different ranges of temperatures and external magnetic fields the fluxons, under the action of the induced mw current, move in different regimes.

*Results at $T \ll T_c$*

At low temperatures and in the range of magnetic fields here investigated, the terms $\lambda$ and $\lambda^2/\delta^2$ in Eq. (2.2) can be neglected; furthermore, it is reasonable to assume that the fluxon dynamics is governed by the restoring-pinning force. In this case $\lambda \ll \lambda_c \ll \delta_f$ and one obtains

$$\tilde{\lambda} \approx \lambda_c (1 - i\lambda_c^2/\delta_f^2), \tag{2.7}$$

and, using Eq.s (2.5) and (2.6),

$$R_S = -\frac{4\pi\omega}{c^2} \text{Im}(\tilde{\lambda}) = \frac{\pi\omega^2 \eta}{c^2} \sqrt{\frac{\phi_0 B}{\pi k_p^3}}. \tag{2.8}$$

So, the $R_S(H)$ curves should be described by the $B^{1/2}$ law. The lines of Fig. 4 are plots of

$$\frac{R_S(H_0)}{R_N} = \frac{R_0}{R_N} + \alpha \sqrt{H_0 - H^*}, \tag{2.9}$$

where for $R_0/R_N$ we have used the experimental values of the normalized residual resistance at $H_0 = 0$, and we have used $\alpha$ and $H^*$ as phenomenological parameters. The best-fit curves have been obtained with $H^* = 280$ Oe for both samples and $\alpha = 2.5 \times 10^{-3}$ for the P#1 sample, $\alpha = 5 \times 10^{-3}$ for the P#2 sample. The $H^*$ value is consistent with the values of $H_{c1}$ reported in the literature for the $MgB_2$ superconductor [2].

We remark that, though the magnetic field dependence of $R_S$ agrees with that expected in the pinning limit, at so low temperatures such large values of the parameter $\alpha$ look unreasonable. By using the same experimental procedure, we have performed measurements of surface resistance in other superconductors. The results of the investigation carried out in YBa$_2$Cu$_3$O$_7$ (YBCO) have shown that up to $H_0$ = 10 kOe no detectable field variations of $R_S$ are observed for $T < 0.7\,T_c$. In order to explore a superconductor with transition temperature and upper critical field smaller than those of YBCO, we have investigated a sample of Ba$_{0.6}$K$_{0.4}$BiO$_3$ (BKBO), $T_c$ = 31 K; in this sample, at $T$ = 4.2 K, we have found that a magnetic field of 10 kOe induces a variation of $R_S$ about ten times less than that observed in MgB$_2$.

The enhanced field variation of $R_S$ at low temperatures cannot be easily justified; on the other hand, it has been shown by several authors that the thermal and transport proprieties of MgB$_2$ are strongly affected by the magnetic field [9-13]. The enhanced field dependence of the properties of MgB$_2$ has been ascribed to the double-gap structure of this compound, with a larger gap, associated with the two-dimensional σ band, and a smaller gap, associated with the three-dimensional π band, which is rapidly suppressed by the applied magnetic field [14-16]. We think that the enhanced field-induced variation of $R_S$ could be due to the unusual structure of vortices related to the different field dependence of the two gaps [14]. As one can see from Eq. (2.8), the value of $R_S$ depends on $\eta$ and $k_p$; it is not easy to understand how the unusual vortex structure influences such parameters; however, it is reasonable to hypothesize that it could affect both the vortex-vortex and the pinning-vortex interactions.

*Results at Temperatures Close to $T_c$*

At temperatures close to $T_c$, where the pinning effects are weak, the fluxons should move in the flux-flow regime, in this case $\delta_v^2 \sim i\delta_f^2/2$; furthermore, in Eq. (2.2) the terms $\lambda$ and $\lambda^2/\delta^2$ cannot be neglected. So, a simple expression for $R_S(H_0)$ cannot be written. Assuming for the viscous-drag coefficient the expression $\eta = \phi_0 B_{c2}(T)/c^2\rho_n$, the only two parameters necessary to perform a comparison between the experimental and the expected $R_S(H_0)$ curves are just the $H_{c2}(T)$ values, which we have deduced from the experimental data, and the ratio $\lambda_0/\delta_0$. It is easy to see that $\lambda_0/\delta_0 = (\omega\tau/2)^{1/2}$, where $\tau$ is the scattering time of the normal electrons. Values of $\tau$ reported in the literature for ceramic MgB$_2$ are $\tau \sim 10^{-13}$ s [8]; so, for these samples at the microwave frequencies the ratio $\lambda_0/\delta_0$ should be of the order of $10^{-2}$. The dashed line of Fig. 5 (b) shows the expected curve obtained using Eqs. (2.1) - (2.6) with $\lambda_0/\delta_0 = 10^{-2}$ and $H_{c2}$ = 6 kOe. As one can see, for $H_0 \ll H_{c2}$, the experimental curve $R_S(H_0)$ varies faster than the expected one; on the contrary, for $H_0 \sim H_{c2}$, the experimental curve varies more slowly than the theoretical one. We have calculated curves of $R_S(H_0)$ using for $\lambda_0/\delta_0$ values ranging from $10^{-1}$ to $10^{-3}$ and for $H_{c2}(T)$ the values deduced from the experimental data, letting them vary within the experimental accuracy. The results have shown that the model, in the present form, does not account for the experimental data. However, the results can be justified quite well by taking into due account the upper-critical-field anisotropy.

For superconducting materials in which the coherence length is much larger than the periodicity of the crystal lattice, such as the MgB$_2$ compound, one can reasonably assume that the angular dependence of the critical field follows the anisotropic Ginsburg-Landau (AGL) theory:

$$H_{c2}(\theta) = \frac{H_{c2}^{\perp c}}{\sqrt{\gamma^2 \cos^2(\theta) + \sin^2(\theta)}}, \tag{2.10}$$

where $\gamma = H_{c2}^{\perp c} / H_{c2}^{\| c}$ is the anisotropy factor and $\theta$ is the angle between the c-axis of crystallites and the external magnetic field.

The values of $H_{c2}$ and its anisotropy factor reported in the literature for MgB$_2$ strongly depend on the kind of sample investigated [3-7, 13]; furthermore, a temperature dependence of $\gamma$ has been highlighted by several authors [3-6]. The AGL theory cannot surely justify the temperature dependence of the anisotropy factor. Nevertheless, it has been shown that data obtained by different techniques can be well justified using Eq. (2.10) in a large range of temperatures, magnetic fields and angles [3, 6, 7]. In order to quantitatively account for our experimental data we have assumed $H_{c2}(\theta)$ of Eq. (2.10). To take into account the effect of the $H_{c2}$ anisotropy in the field dependence of the surface resistance, we have to average the expected $R_S(H_0, H_{c2}(\theta))$ curves over a suitable distribution of the crystallite orientations. We have assumed that our samples consist of crystallites with a random orientation of the c-axes with respect to the applied magnetic field. On this hypothesis, the distribution function of the grain orientation will be

$$dN(\theta) = N_0 \sin(\theta) d\theta/2, \tag{2.11}$$

where $N_0$ is the total number of crystallites.

The expected $R_S(H_0)$ curves depend on the values of $H_{c2}^{\perp c}(T)$, $\lambda_0/\delta_0$ and $\gamma$. On the other hand, the criterion used for deducing the $H_{c2}(T)$ values reported in Fig. 3 allows to determine the field values at which the whole sample goes into the normal state, i.e., the values of $H_{c2}^{\perp c}$. Therefore, in order to fit the data, we have used for $H_{c2}^{\perp c}(T)$ the values deduced from the experimental results, letting them vary within the experimental uncertainty, and we have taken $\lambda_0/\delta_0$ and $\gamma$ as parameters. However, we have found that the expected results are little sensitive to variations of $\lambda_0/\delta_0$ as long as it takes on values smaller than 0.1. Since higher values of this parameter are not reasonable, it cannot be determined by fitting the data, we have used $\lambda_0/\delta_0 = 10^{-2}$; so, the only free parameter is $\gamma$. By fitting the experimental data we have found that, in the range of temperatures investigated, $\gamma$ does not depend on temperature, for both samples. The best-fit curves have been obtained using different values of $\gamma$ for the two samples. In particular, we have obtained $\gamma = 3 \pm 0.1$ for P#1 and $\gamma = 5 \pm 0.3$ for P#2. The continuous lines of Fig. 4 are the best-fit curves of the experimental data.

The values of the anisotropy factor, reported in the literature for MgB$_2$, strongly depend on the type of sample investigated [3-7, 13]. In randomly oriented powder, it has been found $\gamma = 6 - 9$ at low temperature and $\gamma \approx 3$ close to $T_c$ [4, 5]. The $\gamma$-value we have obtained for P#1 agrees with those reported in powdered [4, 5] as well as in polycrystalline samples, on the contrary for P#2 we have obtained a higher value. So, although the two investigated samples have similar shape and have been investigated by the very same measurement method, we have found different values of $\gamma$. This finding corroborates the idea that the anisotropy of the upper critical field depends on the sample growth method, probably because of the different contaminating impurities.

## 3. Harmonic Emission

It has been widely shown that high-$T_c$ superconductors are characterized by markedly nonlinear properties when exposed to intense *em* fields up to microwave frequencies [19, 20, 27-33, 36, 42, 43]. A suitable method for investigating the nonlinear response consists in the detection of signals at the harmonic frequencies of the driving field [27-36]. Besides the basic point of view, this issue takes on great relevance for the use of superconductors in technological applications [19, 20, 27]. Indeed, the nonlinearity is really the main limiting factor for application of superconductors in passive microwave devices; on the other hand, it can be conveniently exploited for assembling active mw devices. For this reasons, it is of great importance recognizing the mechanisms responsible for the nonlinear response and determining the conditions in which the nonlinear effects are important. In the following sections, we report and discuss results of second harmonic (SH) emission by $MgB_2$ ceramic samples. Particular attention will be devoted to the investigation of the properties of the SH signals for applied magnetic fields lower than $H_{c1}$.

### *3.1 Experimental apparatus and samples*

We have investigated the SH emission in several ceramic samples, powdered and bulk, produced by different techniques. However, since for all the samples we have obtained similar results, in this paper we report results in three different samples obtained from Alfa-Aesar powder. The P#α sample consists of ≈ 5 mg of the pristine powder; the second, P#c, has been obtained by further crushing the powder; the third, P#d has been obtained by dispersing the crushed powder in polystyrene, with a ratio of 1:10 in volume.

    The sample is located inside a bimodal rectangular cavity resonating at the two angular frequencies $\omega$ and $2\omega$, in a region where the magnetic fields **H**($\omega$) and **H**($2\omega$) are parallel and of maximal intensity. The fundamental frequency $\omega/2\pi$ is ≈ 3 GHz. The $\omega$-mode of the cavity is fed by a pulsed oscillator, with pulse repetition rate of 200 pps and pulse width of 5 µs, giving a maximum peak power of ≈ 50 W. The sample is also exposed to static magnetic fields, $H_0$, which can be varied from 0 to 10 kOe. The harmonic signals are detected by a superheterodyne receiver. Further details of the experimental apparatus are reported in Ref. [32]. The intensity of the SH signal has been measured as a function of the external magnetic field, the temperature and the input power. All the measurements have been performed with **H**($\omega$)||**H**($2\omega$)||**H**$_0$.

*3.2 Experimental results*

Fig. 6 shows the SH signal intensity as a function of the temperature for the P#α sample. The SH emission is significant in the whole range of temperatures investigated and exhibits an enhanced peak at temperatures close to $T_c$. The kink at $T \approx 38$ K is ascribable to the inhomogeneity of the sample; indeed, the same effect has been found in the temperature dependence of the ac susceptibility [44]. The inset shows the temperature dependence of the SH signal in a restrict range of temperatures around the peak, at three different values of the dc magnetic field. As one can see, on increasing $H_0$ the peak position shifts toward lower temperatures.

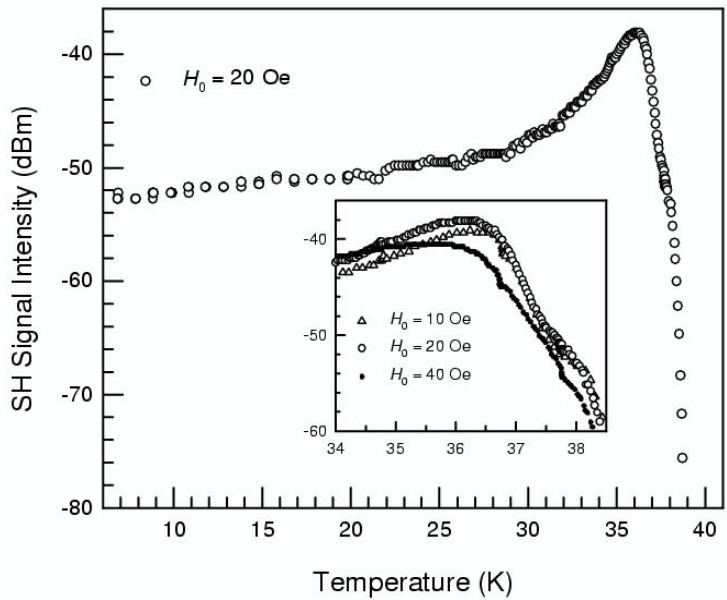

**Figure 6** – SH signal intensity as a function of the temperature for the P#α sample. The inset shows a detail of the temperature dependence of the SH signal, near the peak, at three different values of the dc magnetic field. Input peak power ≈ 30 dBm.

In Fig.7 we report the SH signal intensity as a function of the dc magnetic field for the P#α sample, at $T = 4.2$ K. Full symbols refer to the results obtained in the zero-field-cooled sample on increasing the field for the first time. As expected from symmetry considerations, the SH signal is zero at $H_0 = 0$ (the noise level is ~ -75 dBm); on increasing the field it abruptly increases, exhibits a maximum at ≈ 5 Oe, decreases monotonically up to ~ 40 Oe and then slowly increases until it reaches a value that remains roughly constant up to high fields. Open symbols describe the field dependence of the SH signal observed on decreasing $H_0$, after the first run to high fields. The low-field structure disappears irreversibly after the sample has been exposed to fields higher than ~ 100 Oe; in this case, even at $H_0 = 0$ the intensity of the SH signal takes on roughly the same value as the one measured at high fields.

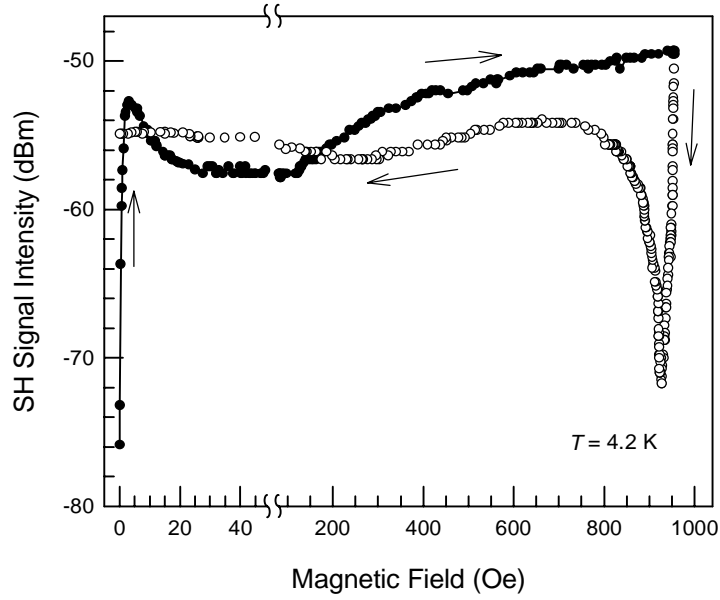

**Figure 7** – SH signal intensity as a function of the dc magnetic field for the P#α sample, at $T = 4.2$ K. Full symbols refer to the results obtained in the zero-field-cooled sample on increasing the field for the first time. Open symbols describe the field dependence of the SH signal on decreasing $H_0$ after the first run to high fields. Input peak power ≈ 27 dBm.

Fig. 8 shows the low-field behaviour of the SH signal, observed in the zero-field-cooled P#α sample by cycling the magnetic field in the range $-H_{max} \div +H_{max}$. The signal shows a hysteretic loop having a butterfly-like shape; on increasing the value of $H_{max}$, the hysteresis is gradually more enhanced and the sharp minima, observed at low fields, move away from each other. The hysteretic loop maintains the same shape in subsequent field runs as long as the value of $H_{max}$ is not changed.

Measurements performed at different input-power levels have shown that, for fixed values of $H_{max}$, the amplitude of the hysteresis increases on decreasing the input power, suggesting that the power dependence of the SH signal is different for increasing and decreasing fields. Fig. 9 shows the intensity of the SH signal as a function of the input power level, at $H_0 = 20$ Oe. Full symbols show the results obtained at $H_0 = 20$ Oe reached at increasing fields; open symbols those obtained at $H_0 = 20$ Oe, reached at decreasing fields after that $H_{max}$ had reached the value of 35 Oe. The distances between full and open symbols, at fixed input power levels, indicate the amplitude of the hysteresis at $H_0 = 20$ Oe.

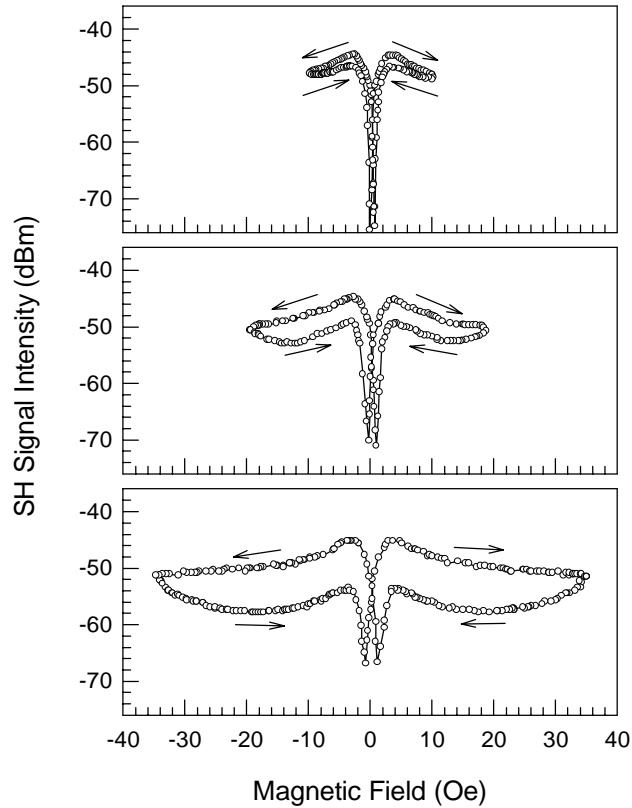

**Figure 8** – Low-field behaviour of the SH signal, observed in the zero-field-cooled P#α sample by cycling the magnetic field in range $-H_{max} \div +H_{max}$. $T = 4.2$ K; input peak power ≈ 30 dBm. The three curves refer to different values of $H_{max}$.

The results reported in Figs. 6-9 refer to the P#α sample. The same measurements have been performed in the other samples, obtained from the P#α sample as described in Section 3.1. The main peculiarities of the SH signals observed in the P#c and P#d samples are the same as those obtained in the P#α sample, except for the relative intensity of the near-$T_c$ peak and the low-$T$ signal. This finding is shown in Fig. 10 where we report the SH signal intensity as a function of the temperature in the three different samples, at $H_0 = 10$ Oe. For the sake of clearness, the three curves have been shifted one with respect to the other, as shown in the figure.

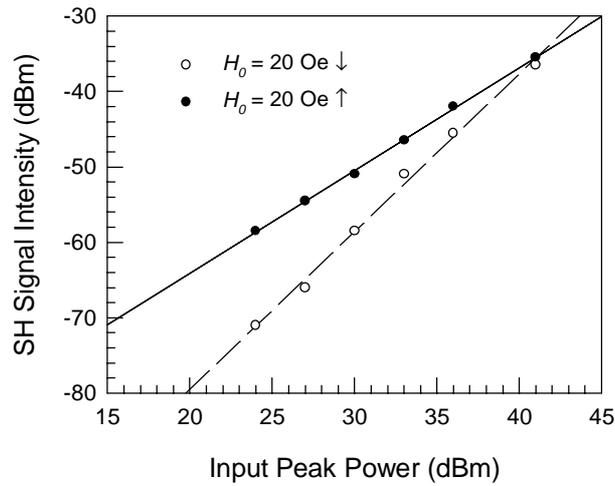

**Figure 9** – SH signal intensity as a function of the input power level in P#α. Full symbols are the results obtained at $H_0 = 20$ Oe reached at increasing fields; open symbols are those obtained at $H_0 = 20$ Oe, reached at decreasing fields.

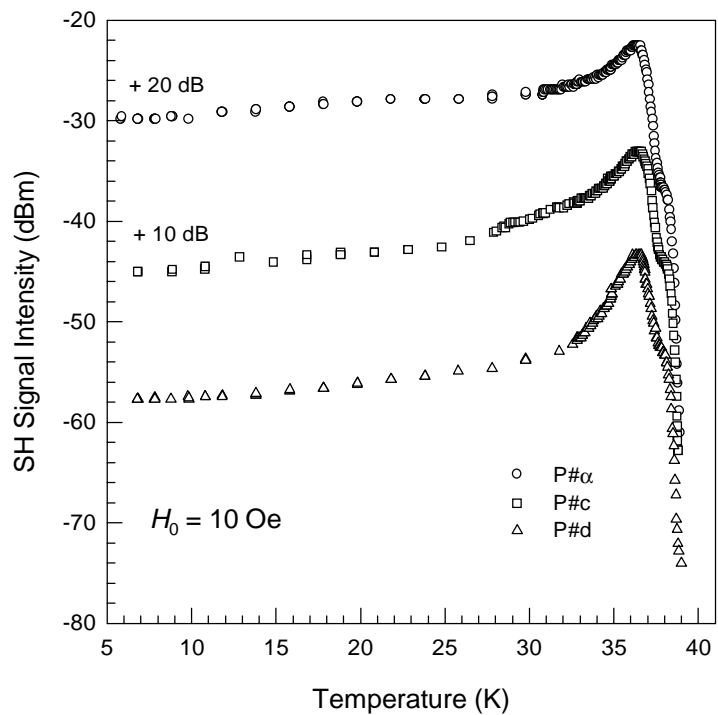

**Figure 10** – SH signal intensity as a function of the temperature in the three different samples. $H_0 = 10$ Oe; input peak power ≈ 30 dBm. The curves relative to P#c and P#α samples have been shifted upwards of 10 dB and 20 dB, respectively.

*3.3 Discussion*

The nonlinear response of high-$T_c$ superconductors to electromagnetic fields has been extensively investigated [27-36, 42, 43]. The nonlinear response detected at low temperatures has been ascribed to extrinsic properties of the superconductors such as impurities, weak links [29, 31, 33-36] or flux-line motion [27, 28]. On the contrary, nonlinearity at temperatures close to $T_c$ is related to intrinsic properties of the superconducting state [30-33, 36]; in particular, it has been ascribed to modulation of the order parameter induced by the *em* field. Our results on $MgB_2$ suggest that, also in this class of superconductors, different mechanisms come into play in the two ranges of temperature. In the following, we discuss the low-*T* and high-*T* behaviour of SH emission by considering models previously reported in the literature.

*Low-Temperature Behaviour*

Harmonic generation has been thoroughly investigated in both ceramic and crystalline YBCO samples [27-35]. It has been shown that at low fields harmonic emission by high-quality crystals is noticeable only at temperatures near $T_c$ [30, 32, 33], while in ceramic samples it is significant at low temperatures as well [29, 34]. This finding has suggested that the low-field and low-*T* harmonic signals are due to nonlinear processes occurring in weak links.

In the presence of weak links, two nonlinear processes may come into play. Harmonic emission is expected when supercurrents are induced, by the $H_0$ and $H(\omega)$ fields, in loops containing Josephson junctions (JJ). In this case, the harmonic emission is strictly related to the intrinsic nonlinearity of the Josephson current [29, 35]. On the other hand, inter-grain dynamics of Josephson fluxons (JF) in the critical state may give rise to harmonic emission [35, 36].

The peculiarities of the low-*T* SH signal observed in the $MgB_2$ samples cannot be accounted for neither in the framework of the model that assumes supercurrent loops containing JJ nor in that which deals with nonlinear dynamics of JF. Results similar to those shown in Fig. 8 have been obtained in BKBO crystals [36]; they have been justified by supposing that the above-mentioned nonlinear processes involving weak links come into play simultaneously. We have shown that the combined effect of the $2\omega$ magnetization arising from the supercurrent loops containing JJ and that arising from the dynamics of JF can justify the peculiarities of the hysteretic behaviour of the low-field SH signal [36]. According to what discussed in Ref.[36], we suggest that also in ceramic $MgB_2$ samples the low-*T* and low-field SH signal arises from nonlinear processes due to the presence of weak links inside the samples. This hypothesis is supported by the irreversible loss of the low-field structure, observed in zero-field-cooled samples, after the samples had been exposed to high fields (see Fig. 7). Indeed, when $H_0$ reaches values higher than $H_{c1}$, Abrikosov fluxons penetrate the grains and the JJ are decoupled by the applied field and/or the trapped flux. In this case, harmonic signals are due to intra-grain fluxon dynamics [28].

Concerning the origin of the SH emission at magnetic fields higher than $H_{c1}$, it has been shown that even harmonics can arise in superconductors in the critical state exposed to high-frequency magnetic fields [28]. In this case, because of the rigidity of the fluxon lattice the superconducting sample operates a rectification process of the ac field. Ciccarello et al. [28] have elaborated a model discussing these effects. It has been shown that a peculiarity of the SH signal of superconductors in the critical state is the presence of enhanced dips in the SH-vs-$H_0$ curves, after the inversion of the magnetic-field-sweep direction, independently of the $H_0$ value at which the inversion is operated. The presence of the enhanced dip of Fig. 7 at the extreme of the plot, in which the magnetic field reverses its direction, shows that also in $MgB_2$ after the sample has been exposed to high fields SH emission is due to the intra-grain fluxon dynamics.

*Near-$T_c$ Behaviour*

The enhanced nonlinear emission observed at temperatures close to $T_c$ cannot be accounted for by mechanisms related by the presence of weak links. Indeed, the Josephson-critical current decreases monotonically on increasing the temperature; on the other hand, near $T_c$ no critical state, either for of Josephson or Abrikosov fluxons, can be reasonably hypothesized. So, a monotonic decrease of the SH-signal intensity it expected.

The harmonic emission at temperatures close to $T_c$ has been ascribed to the time variation of the order parameter induced by the *em* field [30-33, 36]. In particular, the features of second- and third-harmonic signals detected in high-$T_c$ superconductors have been accounted for in the framework of the two-fluid model with the additional hypothesis that the *em* field, which penetrates in the surface layers of the sample, weakly perturbs the partial concentrations of the normal and condensate fluids [30, 32, 36]. According to this model, it is expected that on increasing the external magnetic field and/or the input power level the near-$T_c$ peak broadens and shifts toward lower temperatures [30, 32]. The experimental results reported in the inset of Fig. 6 agree with the expected ones, suggesting that the near-$T_c$ peak detected in $MgB_2$ originates from this mechanism. However, as one can see from Figs. 6 and 10, the peak is merged with the tail of the low-$T$ signal and its characteristics could be affected by this contribution.

In order to corroborate the hypothesis that the low-$T$ and high-$T$ SH signals originate from different mechanisms, the first one from processes involving weak links and the second one from the modulation of the order parameter, we have performed measurements in the three different samples (see Fig.10). As an effect of the grinding of sample P#α, by which sample P#c was obtained, a variation of the number of weak links is expected, with a consequent variation of the low-$T$ SH signal intensity. On the other hand, the effective surface of P#c could be larger than that of P#α and, consequently, an increase of the high-$T$ signal is expected. Actually, from Fig. 10 one can see that the ratio of the intensity of the near-$T_c$ peak and the signal at $T = 5$ K is about 6 dB for P#α and 12 dB for P#c. The dispersion of the grains of the P#c in polystyrene, by which P#d was obtained, has been performed to reduce the number of weak links, maintaining unchanged the effective surface of the sample. A comparison between the SH-*vs*-*T* curves of the samples P#c and P#d of Fig. 10 shows, in effect, that the intensity of the near-$T_c$ peak is the same for the two samples, while a further decrease of the low-*T* signal has be observed after dispersion of the powder in polystyrene. These findings confirm that the low-*T* and near-$T_c$ SH signal have different origin and, in particular, they validate the hypothesis that the low-*T* signal arise from processes involving weak links.

**4. Conclusion**

We have reported experimental results on the microwave response of ceramic $MgB_2$. The response has been investigated in the linear and nonlinear regimes, by measuring the surface impedance and the second-harmonic emission, respectively. The surface resistance and the SH signal have been investigated as function of the temperature and the external magnetic field.

The field-induced variations of the surface resistance have been measured with the aim of investigating the dynamics of fluxons in different ranges of temperature. We have shown that the magnetic field strongly affects the microwave surface resistance of $MgB_2$ even at low temperatures and relatively low values of the magnetic field, where the pinning effects should hinder the energy losses. Although the field dependence of the surface resistance at low temperatures follows the law expected in the pinning limit, it is more enhanced than in other superconductors. We have suggested that this enhanced field dependence of $R_S$ could be due to the unusual structure of vortices related to the different field dependence of the two energy gaps in $MgB_2$. The results at temperatures close to $T_c$ have been quantitatively justified in the framework of the Coffey and Clem model, with fluxons moving in the flux-flow regime, taking into account the anisotropy of the upper critical field. Our results have shown that, at least in the temperature range of about 3 K below $T_c$, the upper-critical-field anisotropy follows the AGL theory. The value of the anisotropy factor of the Alfa-Aesar-powder sample agrees with those reported in the literature for randomly oriented powder of $MgB_2$. However, although the two investigated samples have similar shape, their anisotropy factors are different, corroborating the idea, commonly reported in the literature, that the values of the characteristic parameters of the $MgB_2$ compound depend on the preparation method of the samples.

The results of the second-harmonic emission have shown that, similarly to what occurs in other high-$T_c$ superconductors, several mechanisms are responsible for the nonlinear microwave response of $MgB_2$; their effectiveness depends on temperature and intensity of the external magnetic field. The results obtained at low temperature have show that, although the presence of weak links in $MgB_2$ does not noticeably affect the transport properties, it is the main source of nonlinear response at low magnetic fields and low temperatures. After exposing the sample to magnetic field higher than $H_{c1}$, the weak links are decoupled and the SH emission originates from the dynamics of fluxons in the critical state. At temperatures close to $T_c$, a further contribution to the harmonic emission is present; it arises from modulation of the order parameter by the microwave field and gives rise to a near-$T_c$ peak in the temperature dependence of the SH signal intensity.

*Acknowledgements*

The authors are very glad to thank E. H. Brandt, M. R. Trunin, I. Ciccarello, for their continuous interest and helpful suggestions, G. Lapis and G. Napoli for technical assistance, N. N. Kolesnikov and M. P. Kulakov for supplying one of the MgB$_2$ samples.